\documentclass[12pt,a4paper]{article} 

\usepackage{graphics,graphicx,placeins,afterpage,makecell,tablefootnote,
multicol,multirow,caption,textcomp,lineno,tabularx,subfigure,booktabs,
courier,collref,setspace,float}
\usepackage[utf8]{inputenc}
\usepackage[english]{babel}
\usepackage[backend=bibtex,style=authoryear,bibstyle=authoryear,maxcitenames=2,maxbibnames=7,useprefix=true]{biblatex}
\DeclareNameAlias{sortname}{last-first}

\begin{document}
\pagestyle{plain}
\pagenumbering{arabic}
 
\begin{center}
{\LARGE{Spatial Variations in Titan’s Atmospheric Temperature:
      ALMA and \textit{Cassini} Comparisons from 2012 to 2015}}
\end{center}
\bigskip

\begin{center}
Authors: Alexander E. Thelen$^a$\footnote{Corresponding
  author. (A. E. Thelen) Email
  address: athelen@nmsu.edu. Postal Address:
  Department of Astronomy, New Mexico State University, PO BOX 30001,
  MSC 4500, Las Cruces, NM 88003-8001.}, C. A. Nixon$^b$,
N. J. Chanover$^a$, E. M. Molter$^c$, M. A. Cordiner$^{b,d}$,
R. K. Achterberg$^{b,e}$, J. Serigano$^f$, P. G. J. Irwin$^g$, N. Teanby$^h$, S. B. Charnley$^b$
\bigskip

{\footnotesize{$^a$New Mexico State University, $^b$NASA Goddard Space Flight Center,
$^c$University of California, Berkeley, $^d$Catholic University of
America, $^e$University of Maryland, $^f$Johns Hopkins University,
$^g$University of Oxford, $^h$University of Bristol}}
\end{center}
\bigskip


\begin{abstract}
Submillimeter emission lines of carbon monoxide (CO) in Titan's atmosphere
provide excellent probes of atmospheric temperature due to the
molecule's long chemical lifetime and stable, well constrained volume
mixing ratio. Here we present the analysis of 4 datasets obtained with
the Atacama Large Millimeter/Submillimeter Array (ALMA) in 2012, 2013, 2014, and
2015 that contain strong CO rotational transitions. Utilizing ALMA's
high spatial resolution in the 2012, 2014, and 2015
observations, we extract spectra from 3 separate regions on Titan's
disk using datasets with beam sizes ranging from $0.35\times0.28''$ to
$0.39\times0.34''$. Temperature profiles retrieved by the NEMESIS radiative transfer
code are compared to \textit{Cassini} Composite Infrared Spectrometer
(CIRS) and radio occultation science results from similar latitude
regions. Disk-averaged temperature profiles stay relatively constant
from year to year, while small seasonal variations in atmospheric temperature
are present from 2012--2015 in the stratosphere and mesosphere
($\sim$100-500 km) of spatially resolved regions. We measure the
stratopause (320 km) to increase in temperature by 5 K in northern
latitudes from 2012--2015, while temperatures rise throughout the
stratosphere at lower latitudes. We observe
generally cooler temperatures in the lower stratosphere ($\sim$100 km)
than those obtained through \textit{Cassini} radio occultation measurements, with the notable
exception of warming in the northern latitudes and the absence of
previous instabilities; both of these results are indicators that
Titan's lower atmosphere responds to seasonal effects, particularly at
higher latitudes. While retrieved temperature profiles cover a
range of latitudes in these observations, deviations from
CIRS nadir maps and radio occultation measurements convolved with the
ALMA beam-footprint are not found to be statistically significant, and
discrepancies are often found to be less than 5 K throughout the
atmosphere. ALMA's excellent sensitivity in the lower stratosphere (60--300 km)
provides a highly complementary dataset to contemporary CIRS and
radio science observations, including altitude regions where both of
those measurement sets contain
large uncertainties. The
demonstrated utility of CO emission lines in the submillimeter as
a tracer of Titan's atmospheric temperature lays the groundwork for
future studies of other molecular species --
particularly those that exhibit strong polar abundance enhancements or are
pressure-broadened in the lower atmosphere, as temperature profiles
are found to consistently vary with latitude in all three years by up
to 15 K. 
\end{abstract}
\bigskip

\begin{center}
\end{center}
\bigskip

\section{Introduction} \label{introduction}
Titan's atmospheric temperature, strongly influenced by solar
heating, photochemistry, cloud and haze formation, and Hadley-type
circulation, has been shown to have large spatial and temporal
variations in the lower and middle atmosphere ($\leq$600 km). Many measurements
of Titan's atmospheric temperature and dynamics have been made in the previous
decades from ground- and space-based facilities, including the
Submillimeter Array, \textit{Voyager 1}, and
\textit{Cassini} orbiter observations, through radio
occultations, ultraviolet, infrared, near-IR, heterodyne, and submillimeter spectroscopy (see
reviews in \cite{flasar_14}, and \cite{griffith_14}). The \textit{Cassini} spacecraft, in particular,
has provided unprecedented measurements of atmospheric temperature in
Titan's troposphere and stratosphere through modeling CH$_4$ vibrational-rotational
emission in the IR and \textit{in
  situ} measurements by the Huygens Atmospheric Structure Instrument
(HASI) \parencite{fulchignoni_05,
  flasar_05, vinatier_07, dekok_07, achterberg_08,
  teanby_10a, teanby_10b, achterberg_11, teanby_12,
  vinatier_15, coustenis_16}. Additionally, temperatures in the lower
atmosphere have been obtained through \textit{Cassini} radio
occultations, where spacecraft signals are refracted by Titan's
atmosphere during transmission to Earth \parencite{schinder_11, schinder_12}. These datasets have constrained
temperatures to $\leq$1 K over many close flybys of Titan starting in
2004. Throughout \textit{Cassini's} extended tour of the Saturnian system, close to half of a full
seasonal cycle (29.5 years) has been observed on Titan; however,
\textit{Cassini's} finale in 2017 will greatly limit further studies of Titan's
atmosphere with such exceptional resolution and cadence. 

In addition to commonly used atmospheric temperature diagnostics such
as thermal emission and modeling CH$_4$ bands in the IR -- whose
abundance is well constrained in Titan's atmosphere
by \textit{Cassini} and \textit{in situ} measurements by the \textit{Huygens}
probe \parencite{niemann_05} -- CO may also be used as a probe for
atmospheric temperature. Previous observations of Titan in the IR and
submillimeter regimes from ground- and space-based facilities
on Earth have provided many constraints on CO abundance throughout the
atmosphere by utilizing \textit{a priori} temperature measurements from
\textit{Cassini} and \textit{Voyager 1}
observations \parencite{gurwell_00, lellouch_03, gurwell_04, rengel_11, courtin_11,
  gurwell_11, debergh_12, teanby_13, rengel_14}. The volume mixing
ratio of CO, found to be approximately 50 ppm, appears to be
extremely stable throughout Titan's atmosphere as found by
observations and photochemical models due to its long photochemical
lifetime, estimated to be upwards of 75 Myr \parencite{yung_84, gurwell_00,
  lellouch_03, krasnopolsky_14, loison_15}. \textcite{serigano_16}
recently retrieved disk-averaged temperature profiles by modeling CO
emission lines present in flux calibration images of Titan obtained
with the Atacama Large Millimeter/Submillimeter Array (ALMA) during 2012 and 2014, and
determined a constant CO volume mixing ratio of 49.6 ppm.

Here we present the first spatially resolved temperature measurements
of Titan obtained using ground-based radio observations. We analyze data
from four short ($\sim$3 minute) ALMA flux calibration observations of Titan to obtain disk-averaged
measurements in atmospheric temperature from 2012--2015, and
independent temperature measurements of three distinct latitude regions ($\sim48^\circ$
N, $20^\circ$ N, and $15^\circ$ S) in 2012, 2014, and
2015. We compare temperature profiles covering altitudes in Titan's lower stratosphere through
mesosphere (50--500 km) to those obtained by Cassini
Composite Infrared Spectrometer (CIRS) nadir mapping observations and through radio
occultations. Through the future combination of ALMA and \textit{Cassini}
datasets, this technique may be used to monitor Titan's atmospheric temperature
and dynamics into northern summer, completing the seasonal cycle
observed by \textit{Cassini} beginning in northern winter in
2004.

\section{Observations} \label{observations}
\indent We obtained public data of Titan from the ALMA
science
archive\footnote{https://almascience.nrao.edu/alma-data/archive}. Titan
is often used as a flux calibrator for ALMA science targets, allowing
us to utilize frequent (almost daily for $\sim$7 months of the year)
observations over the duration of ALMA's
lifespan for this study. Though many additional datasets
containing short integration times on Titan exist in the ALMA science
archive, we employed only those observations with beam sizes roughly one
third of Titan's angular diameter -- $\sim$1$''$ including the moon's
solid body and atmosphere. During early cycles, the spatial resolution was often
\textgreater$0.20''$ due to fewer available antennae in the array. High spatial resolution data from
2013 are particularly limited, as the ALMA array was
undergoing construction and commissioning during this period,
resulting in few useful observations. Observational parameters are
given in Table \ref{tab:obs} for the datasets used in this
study. 


We completed data reduction in a fashion similar to previous ALMA
studies of Titan \parencite{cordiner_14,
  cordiner_15, serigano_16, molter_16, palmer_17}. We processed data using modified versions of the scripts provided
by the Joint ALMA Observatory, which often flag out strong lines in Titan's
atmosphere (such as CO). We re-ran these
scripts in the NRAO's CASA software package 4.7.0 to correctly include
CO lines, and executed standard protocols such as flagging terrestrial
lines and shadowed data, bandpass and gain calibration. Imaging was
completed using the CASA \texttt{clean} task. Deconvolution of the
ALMA point-spread function was performed by using the H{\"o}gbom
algorithm, with natural visibility weighting and an flux threshold of
twice the expected RMS noise (typically on order 15 mJy). The image
pixel sizes were set to 0.03$''\times$0.03$''$. Image spectral
coordinates were Doppler-shifted to Titan's rest frame using
Topocentric radial velocities from JPL Horizons ephemerides. 

\begin{figure}[H]
\subfigure{\includegraphics[width=6cm]{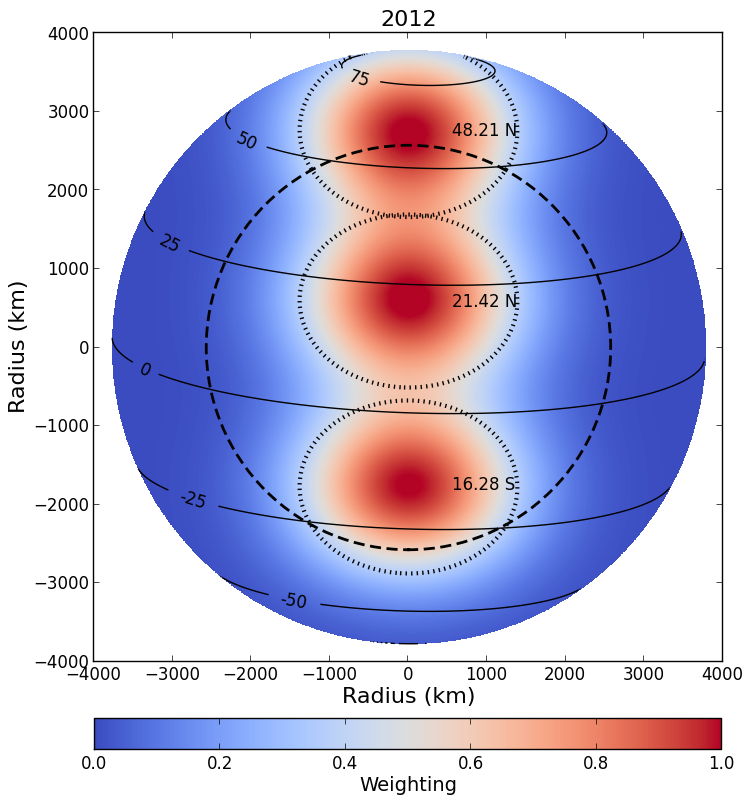}}
\subfigure{\includegraphics[width=6.1cm]{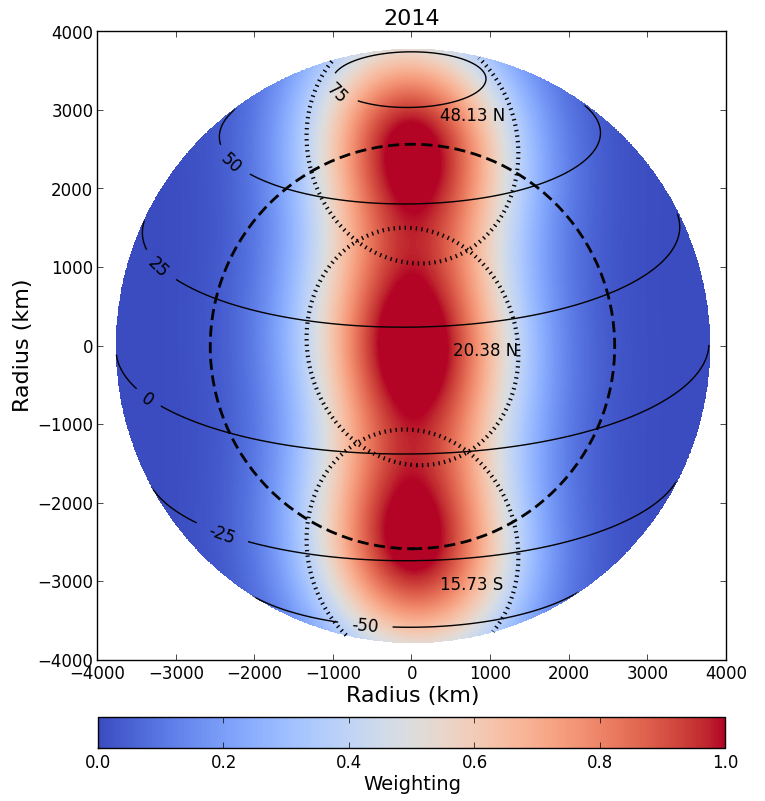}}
\subfigure{\includegraphics[width=6.05cm]{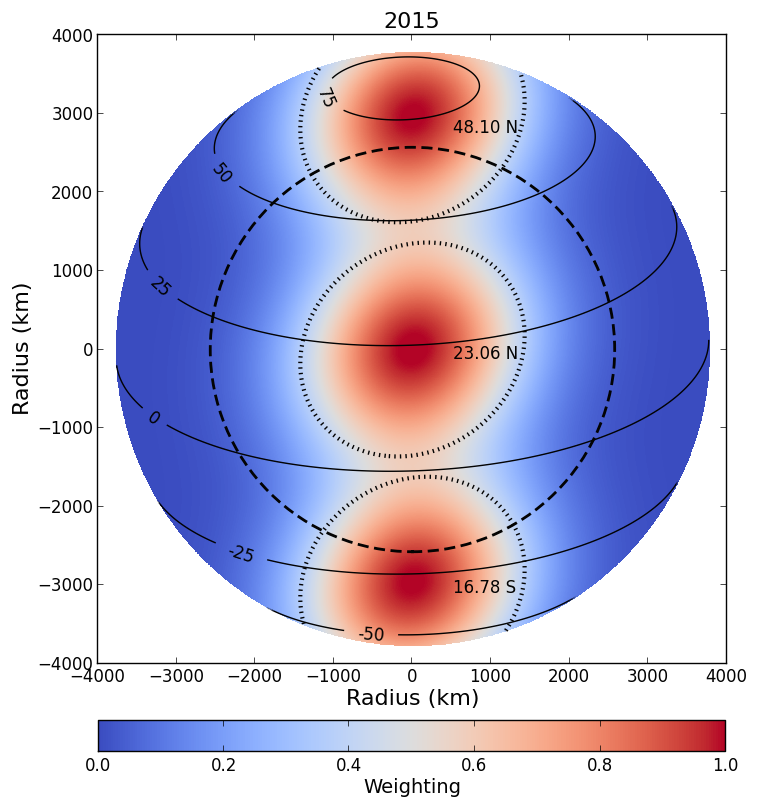}}
\caption{Representative spatial regions for ALMA 2012 (left panel), 2014
    (center), and
    2015 (right) images of Titan. Titan's solid body radius is shown
    as a dashed circle. Solid lines trace the top-of-atmosphere (1200
    km above the surface) latitudes. Color scale represents weighting given to emission angles for
  radiative transfer modeling. Beam-footprints are placed such that
  the mean latitude of each region (printed) does not change by more than
  $5^\circ$ from year to year, while also limiting the flux
  contribution from adjacent regions. Titan's tilt is also accounted for,
  which changed by $\sim12^\circ$ from 2012--2015. Weighting contributions are not coadded, as plotted
  here. The beam FWHM corresponding to each extracted spectrum is
  shown with a dotted ellipse.}
\label{fig:beam}
\end{figure}

We obtained disk-averaged
measurements for each year from 2012--2015 by extracting
flux over Titan's solid disk (2575 km),
plus its extended atmosphere (1200 km) and an additional
2-$\sigma_{PSF}$, where $\sigma_{PSF}$ is the standard deviation of
ALMA's point-spread function, or the FWHM (major axis) of the Gaussian restoring
beam. ALMA configurations in 2012, 2014, and 2015
permitted beam sizes that are smaller than Titan's angular diameter, allowing us to extract independent flux measurements of
multiple, spatially resolved 
regions on Titan's disk. We chose to model spectra from 3 regions
(hereafter referred to as `North',
`Center', and `South') chosen to be as independent as possible (with
minimal beam overlap), while the mean latitudes of extraction regions
are within $5^\circ$ from year to year. These regions are
shown in Fig. \ref{fig:beam}. As such, flux density measurements
from these ``beam-footprints'' are representative of a few latitude decades.

We obtained initial flux density estimates by using the
Butler-JPL-Horizons 2012 Titan flux model in the CASA reduction
scripts, which is expected to be accurate to within 5$\%$ (see ALMA Memo
$\#$594\footnote{https://science.nrao.edu/facilities/alma/aboutALMA/Technology/ALMA$\_$Memo$\_$Series/alma594/memo594.pdf}). This
model is based on previous ground-based submillimeter observations of
Titan, and includes: strong emission lines from trace species in Titan's
atmosphere -- namely CO, HCN, and their respective isotopologues;
collisionally-induced absorption of N$_2$-N$_2$ and N$_2$-CH$_4$;
emission from Titan's surface \parencite{gurwell_00, gurwell_04}.


\section{Spectral Modeling and Results} \label{results}
\indent We converted ALMA spectra to spectral radiance units
(nW/cm$^{2}$/sr/cm$^{-1}$) as described by \textcite{teanby_13} before
performing radiative transfer modeling using the Non-Linear
Optimal Estimator for Multivariate Spectral Analysis (NEMESIS)
software package in line-by-line mode \parencite{irwin_08}. We
obtained spectral line
parameters from the HITRAN 2012 and CDMS databases
\parencite{rothman_13, muller_01} as in \textcite{serigano_16, molter_16}. We calculated collisionally-induced
absorption parameters for N$_2$, CH$_4$, and H$_2$ pairs from the
works of Borysow and Frommhold \parencite*{borysow_86a,borysow_86b,
  borysow_86c, borysow_87}, Borysow \parencite*{borysow_91}, and
Borysow and Tang \parencite*{borysow_93}. We then initialized models
of Titan's atmosphere using N$_2$ and CH$_4$ vertical profiles
from \textcite{niemann_10} and \textcite{teanby_13}, with a constant
CO abundance of 49.6 ppm as found by \textcite{serigano_16}, which is in
agreement with previous CO measurements \parencite{dekok_07, teanby_10b,
  gurwell_11, debergh_12, teanby_13, rengel_14}. Assuming the CO
abundance profile is constant due to its long photochemical
lifetime allows us to fit spectra by only varying vertical temperature profiles, as emission lines of CO are significantly
pressure-broadened in Titan's atmosphere and thus enable temperature
retrievals from a wide range of altitudes. Temperature retrievals
obtained using CO abundances in excess of \textpm$20\%$ of the nominal value (49.6 ppm) resulted in poor spectral
fits or large variations in temperature, as in
\textcite{serigano_16}. Using a non-uniform CO abundance profile with
small variations (\textless5 ppm), such as that found by \textcite{loison_15}, has little effect on spectral fits
and retrieved temperature profiles.

We first model disk-averaged spectra to obtain initial temperature
profiles for each dataset listed in Table \ref{tab:obs}. As emission
from Titan's extended atmosphere results in significant
limb brightening, we generated
spectral models using 37--44 field-of-view averaging points (available
online from [Dataset] \cite{thelen_17a}\footnote{http://dx.doi.org/10.17632/m5pscpthph.1}), as detailed in Appendix A
of \textcite{teanby_13}. As a starting point for our disk-averaged
retrievals, we constructed \textit{a priori} temperature profiles using data constrained by
measurements from \textit{Cassini} CIRS and the \textit{Huygens}
probe \parencite{flasar_05, fulchignoni_05}, as in previous ALMA studies of
Titan \parencite{cordiner_14, cordiner_15, serigano_16,
  molter_16}. Additionally, we also use contemporary disk-averaged
CIRS nadir data (similar to \cite{achterberg_14}), which provide high sensitivity measurements between 0.1--10
mbar. These datasets are detailed in Table \ref{tab:cassini}. Upper atmospheric
temperatures (\textgreater600 km) were held as isothermal at 160
K.

\begin{figure}[H]
\centering
\includegraphics[scale=0.332]{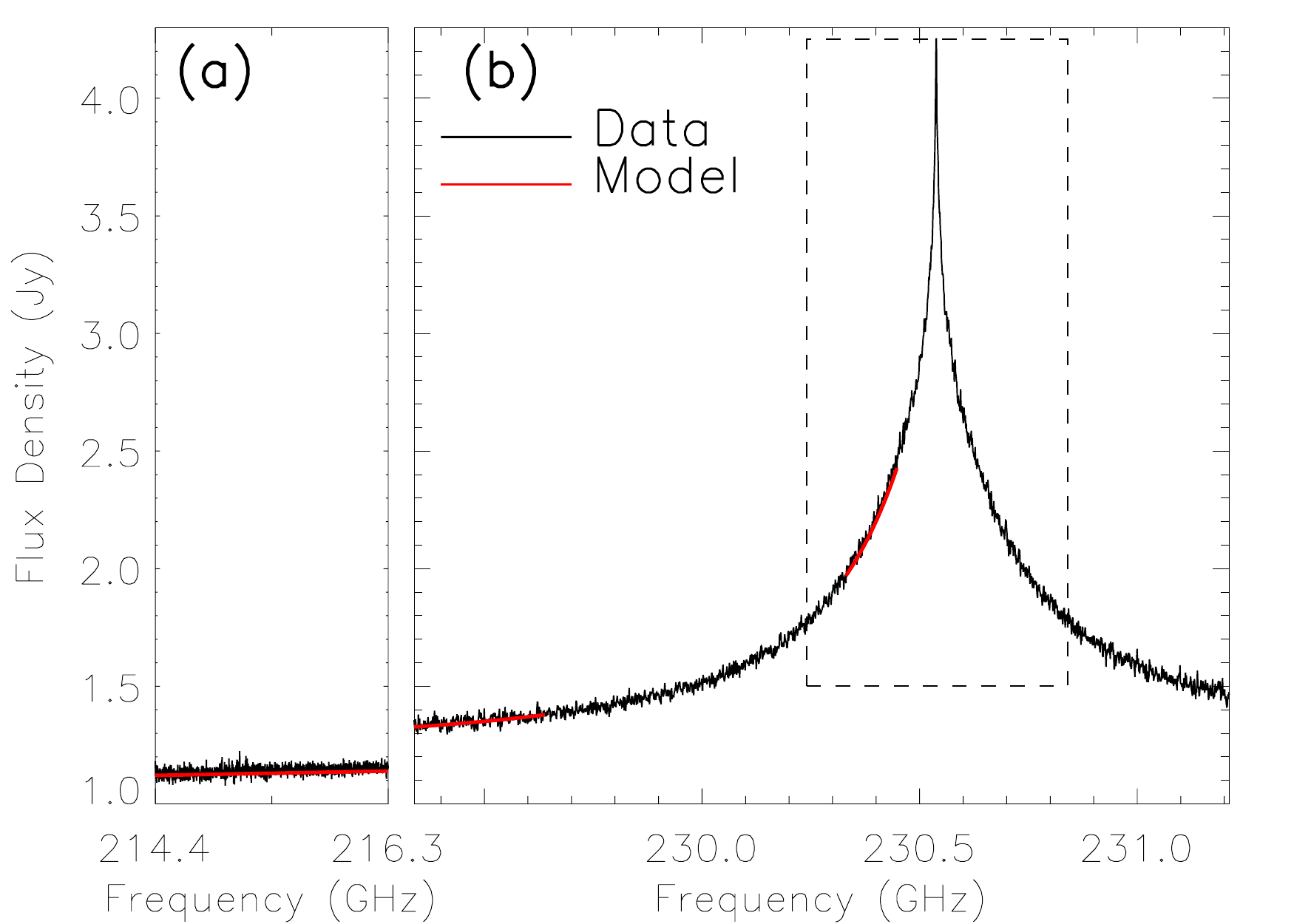}
\includegraphics[scale=0.332]{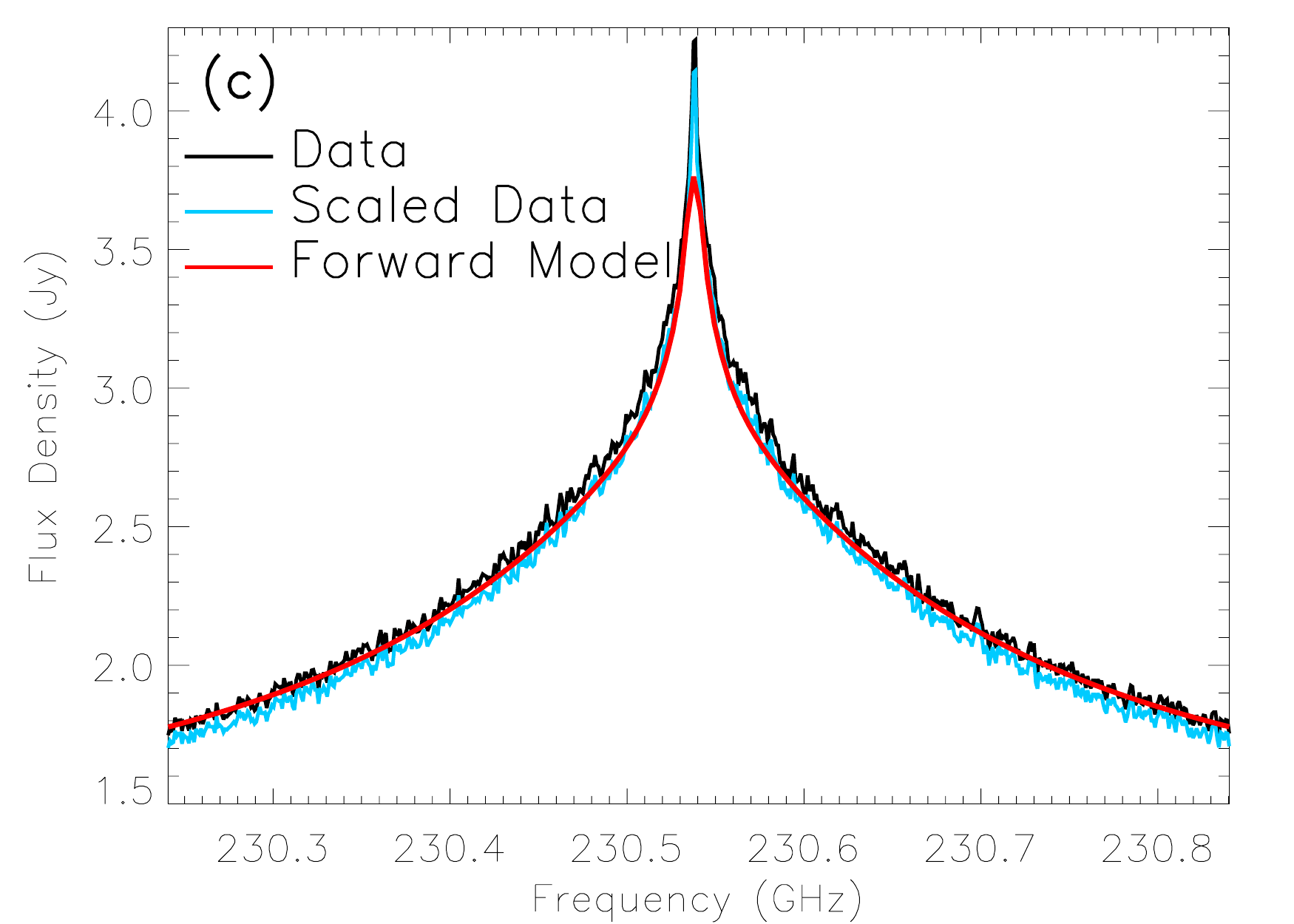}
\includegraphics[scale=0.332]{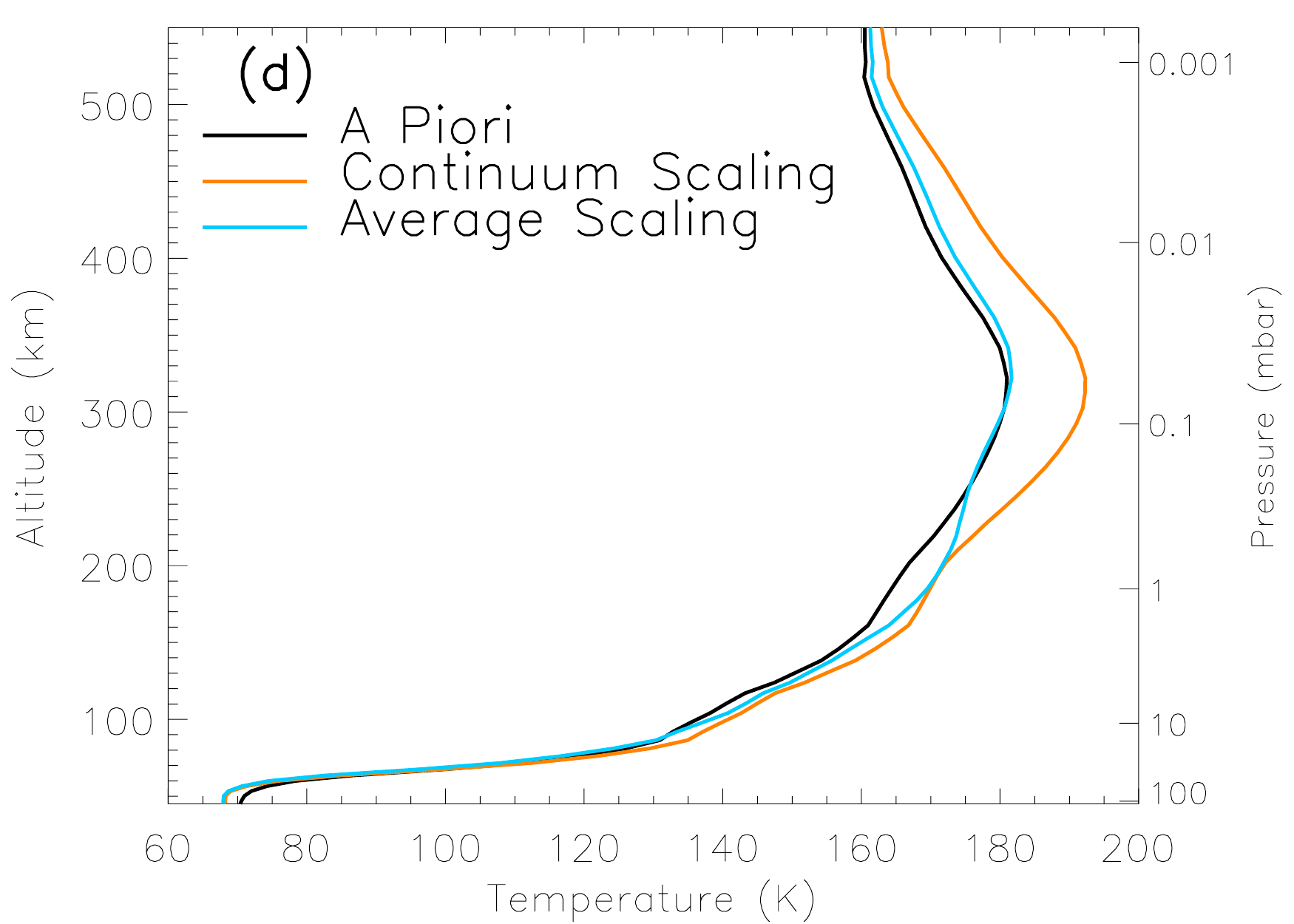}
\caption{(a, b) ALMA spectra (black) from adjacent continuum (a) and CO
    (2-1) line (b) spectral windows from ALMA dataset 2012.1.00317.S
    (Table \ref{tab:obs}). Synthetic spectra (red) show regions
    modeled to determine uniform offsets between data and
    models. The boxed region denotes the spectral range chosen to
    model for temperature retrievals. (c) Close up of boxed region in
    (b). ALMA data (black) is compared to data scaled by a constant
    factor (0.974; teal) obtained by averaging offsets of regions
    shown in (a, b). NEMESIS synthetic spectrum is
  shown before temperature retrieval (red). (d) Temperature retrievals
 for data scaled by the continuum scaling factor (0.990; orange) from
 (a), and average factor (0.974; teal) as shown in (c). The a priori
 temperature profile is plotted in black.}
\label{fig:scale}
\end{figure}

Initial fits of continuum
regions in adjacent spectral windows were greatly improved by utilizing temperature
profiles of Titan's lower atmosphere obtained by \textit{Cassini} radio
occultation science (\cite{schinder_12}; Table
\ref{tab:cassini}). While these observations are not contemporary with
those by ALMA, temperatures near Titan's tropopause (40--60
km, where the radio continuum is formed) are not expected to change
on seasonal timescales, but do vary with latitude \parencite{flasar_81, schinder_12}. We
then multiplied our spectra by a small scaling factor (\textless5$\%$
of the spectral radiance) due to discrepancies
between our model and that of Butler-JPL-Horizons 2012. These
discrepancies may be due to: slight latitudinal troposphere temperature
variations (on the order 5 K) between
the models; minor uncertainties in line broadening or
collisionally-induced absorption parameters; and variations in CO
abundance on the order of 10$\%$. A constant scaling factor was
determined for each spectrum by averaging offsets between measurements
of data and model continuum, CO line wings, and line core regions, to
distinguish between minute flux
calibration issues during the ALMA pipeline and effects caused by
temperature variations in Titan's atmosphere. Synthetic spectra
generated to determine scaling factors for the 2015 CO (2--1) line,
along with a comparison of the CO line forward
model, unscaled, and scaled data are shown in Fig. \ref{fig:scale}
(panels a, b, and c). The effects of these scaling factors on the
retrieved temperature profiles are shown in
Fig. \ref{fig:scale}d. 

We retrieve vertical temperature profiles by allowing NEMESIS to vary temperature
measurements continuously throughout the atmosphere, with 5 K errors
set on the \textit{a priori} temperature profile; combined with a
correlation length of 1.5 scale heights, these errors enable
NEMESIS to adjust temperature profiles enough to obtain excellent
spectral fits, but reduce ill-conditioning (artificial vertical
structure; \cite{irwin_08}). We assume that small scale vertical
structure is a result of NEMESIS fitting noise in the spectrum, and is
not a result of atmospheric dynamics (e.g. gravity waves) -- these
correlation length and \textit{a priori} errors are large enough to
properly constrain the spectral fit, but provide sufficient smoothing
to prevent unrealistic vertical oscillations.



\begin{figure}[H]
\centering
\includegraphics[scale=1.05]{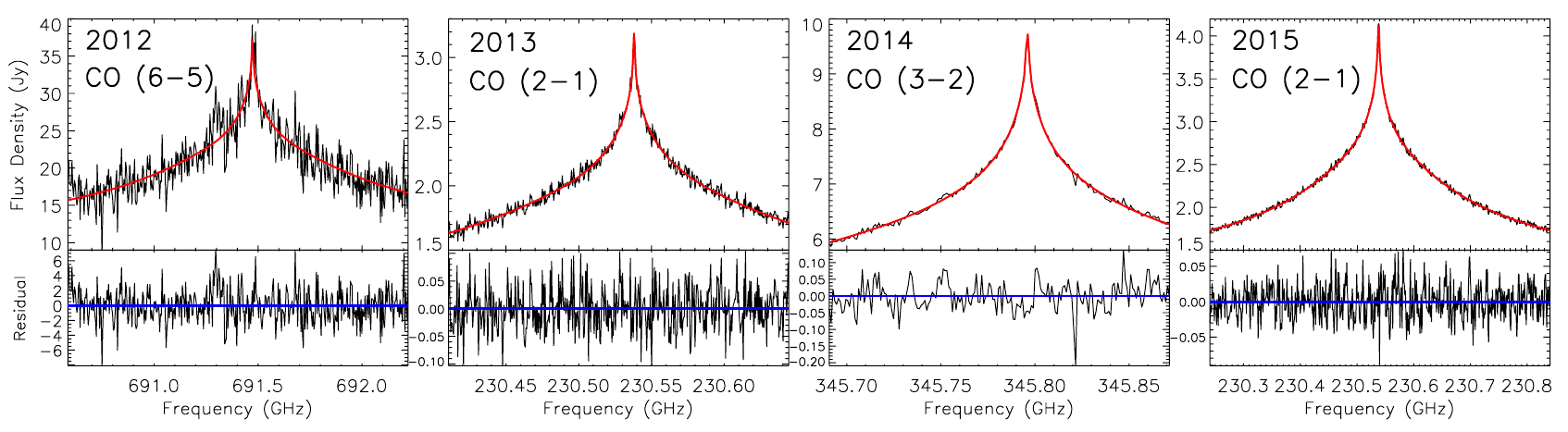}
\caption{ALMA spectra (black) and synthetic spectra generated by
    the NEMESIS radiative transfer code (red) for disk-averaged
    measurements of flux from 2012--2015 datasets. Bottom panels show
    the residual flux after subtracting the model from the
    observed spectrum.}
\label{fig:spec1}
\end{figure}

\begin{figure}[H]
\centering
\includegraphics[scale=0.47]{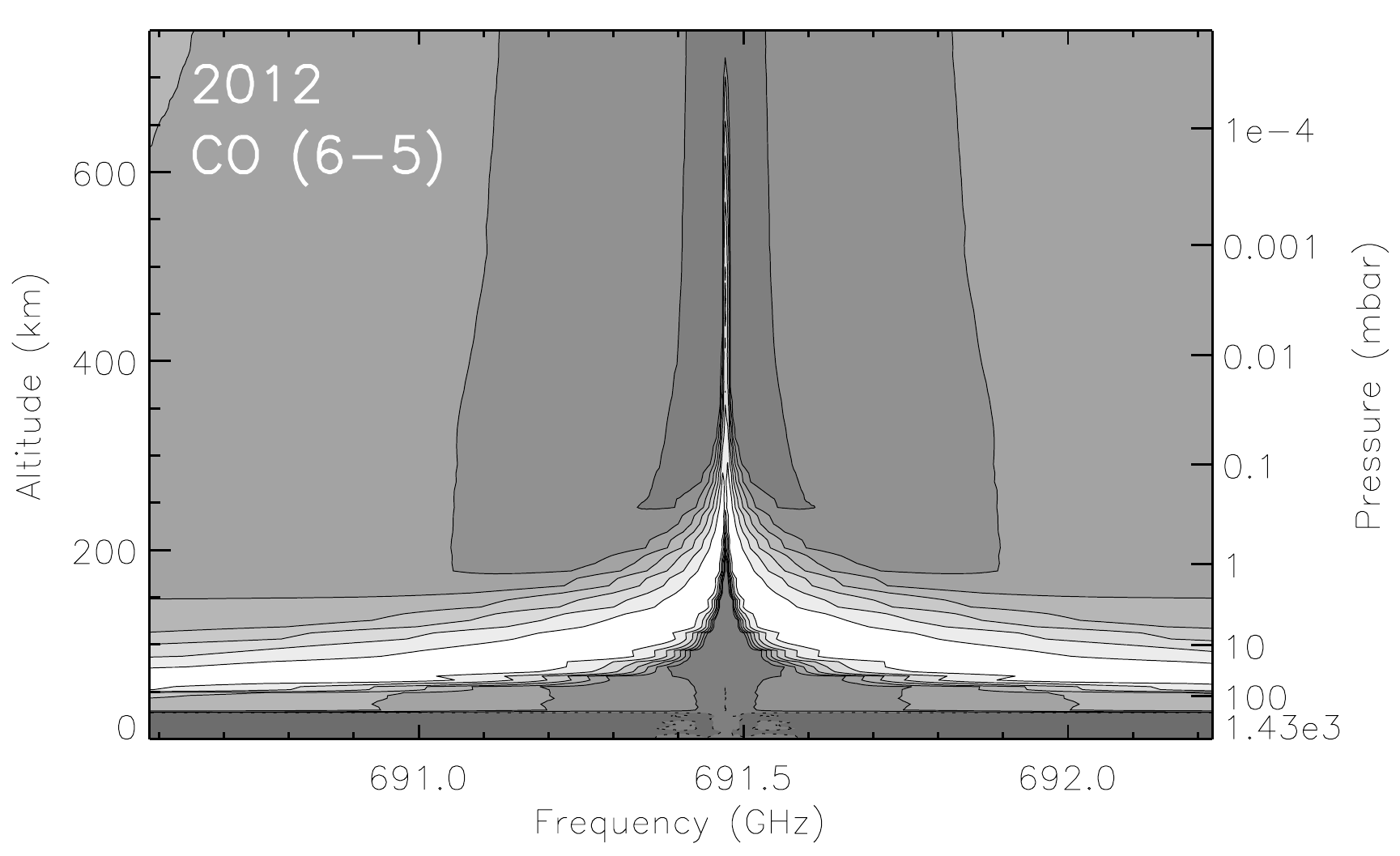}
\includegraphics[scale=0.47]{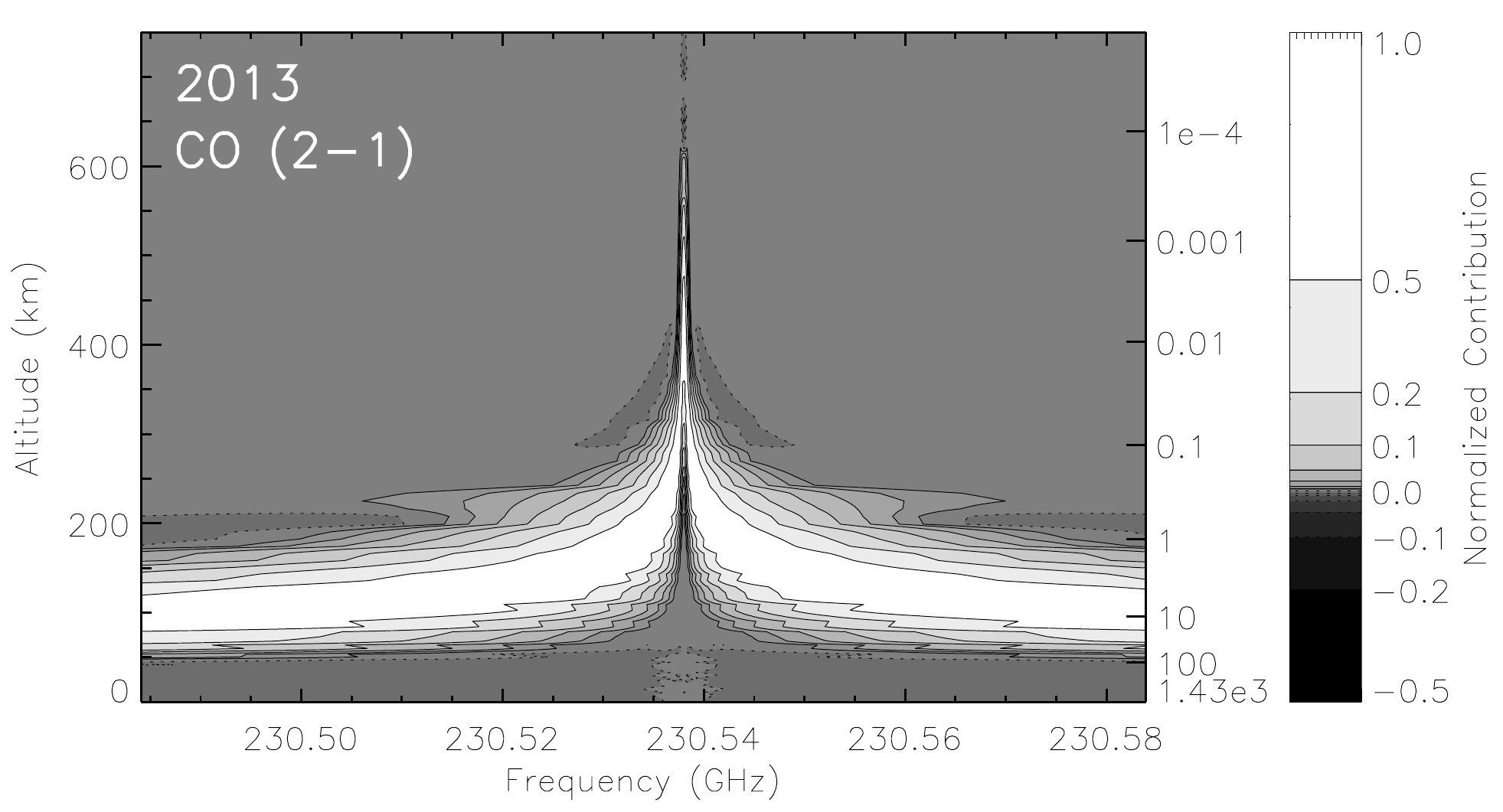}
\includegraphics[scale=0.47]{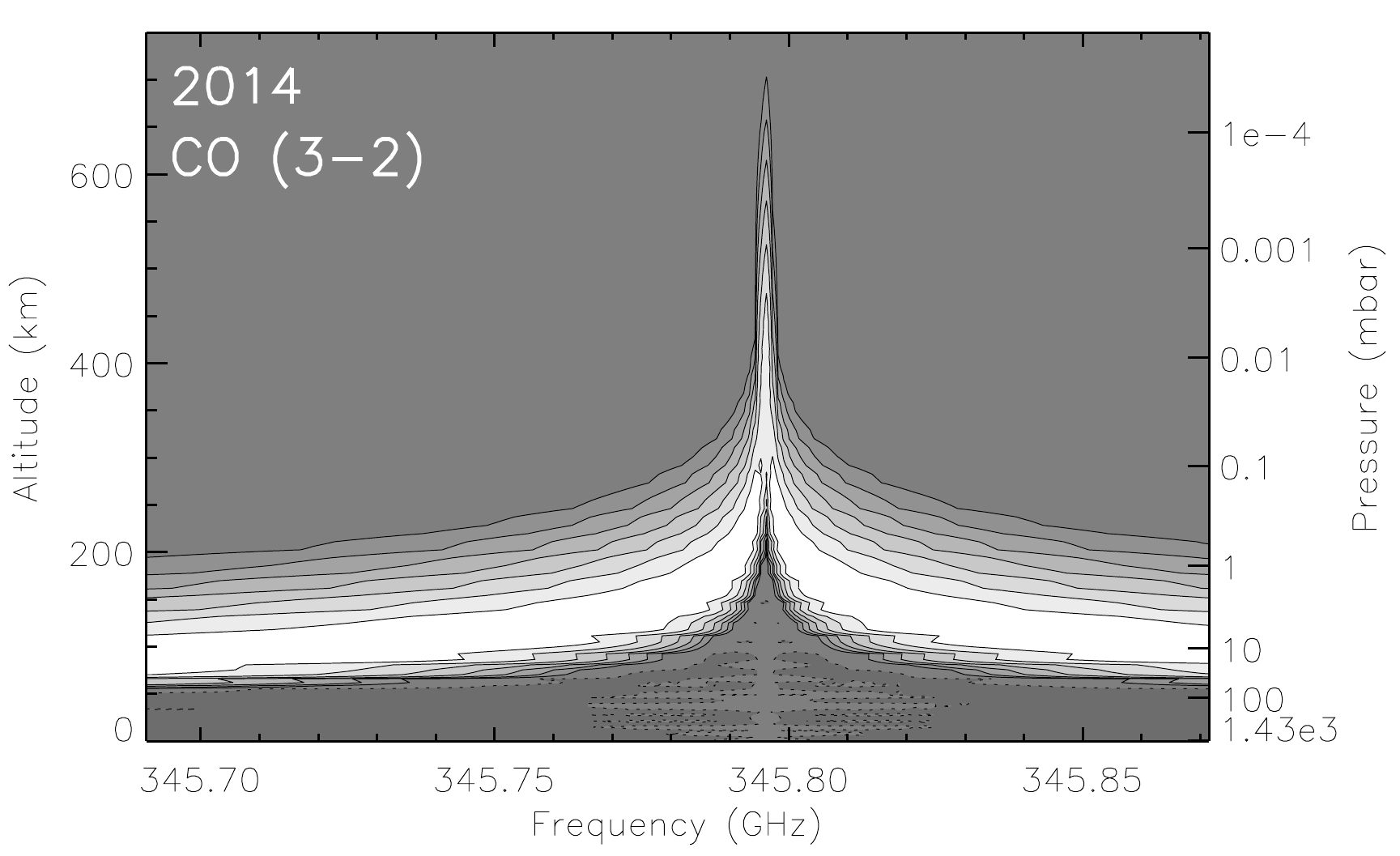}
\includegraphics[scale=0.47]{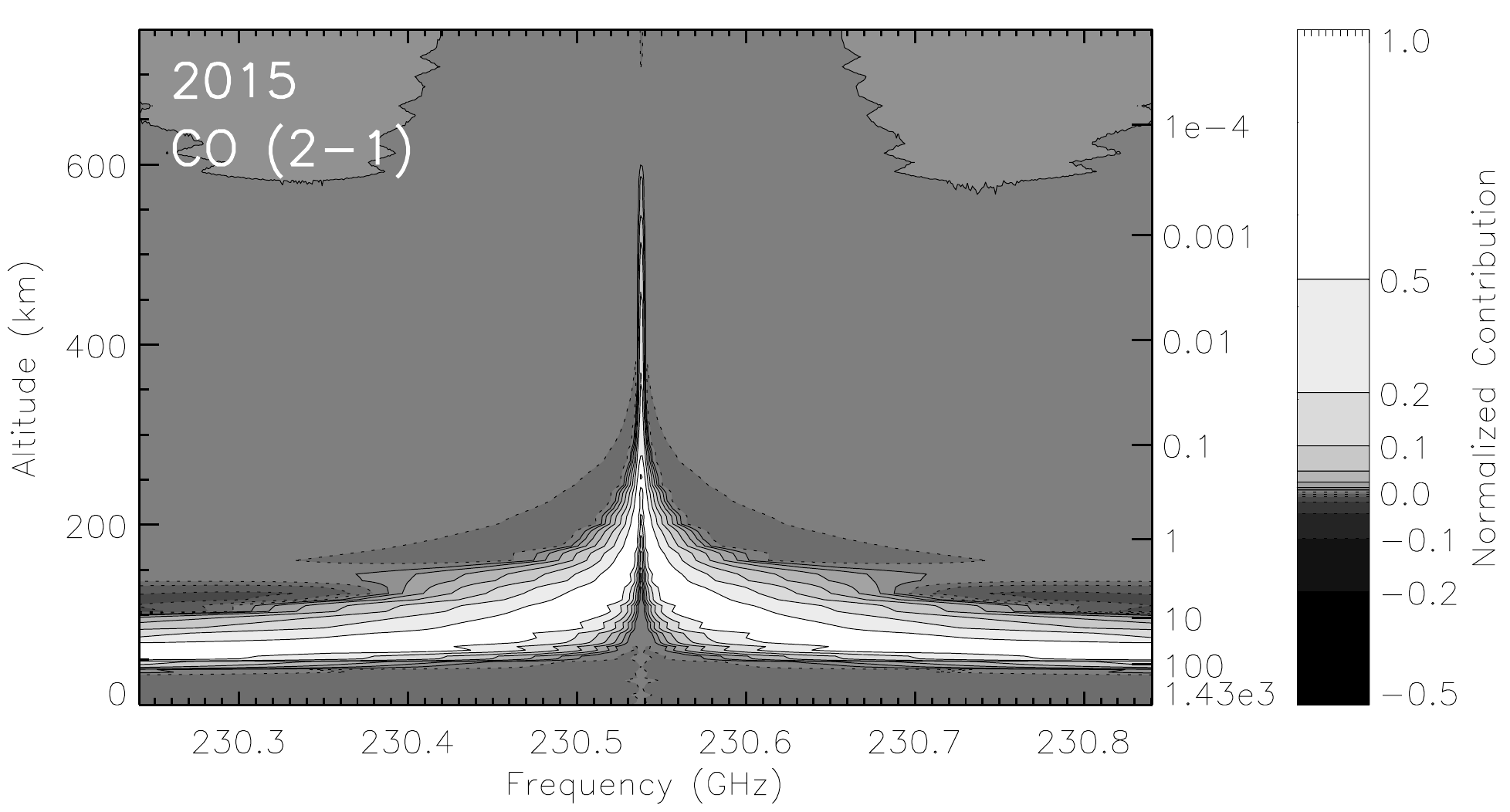}
\caption{Contours of normalized functional derivatives \parencite{irwin_08} of spectral radiance per
    wavenumber with respect to temperature for each disk-averaged spectrum in
    Fig. \ref{fig:spec1}. Contour levels are 0, \textpm0.0046, \textpm0.01, \textpm0.0215,
    \textpm0.046, \textpm0.1, \textpm0.215, and \textpm0.46, as in
    \textcite{molter_16}; levels express CO emission sensitivity to
    temperature at various pressure and altitude values.}
\label{fig:cf}
\end{figure}

\begin{figure}[H]
\centering
\includegraphics[scale=1.2]{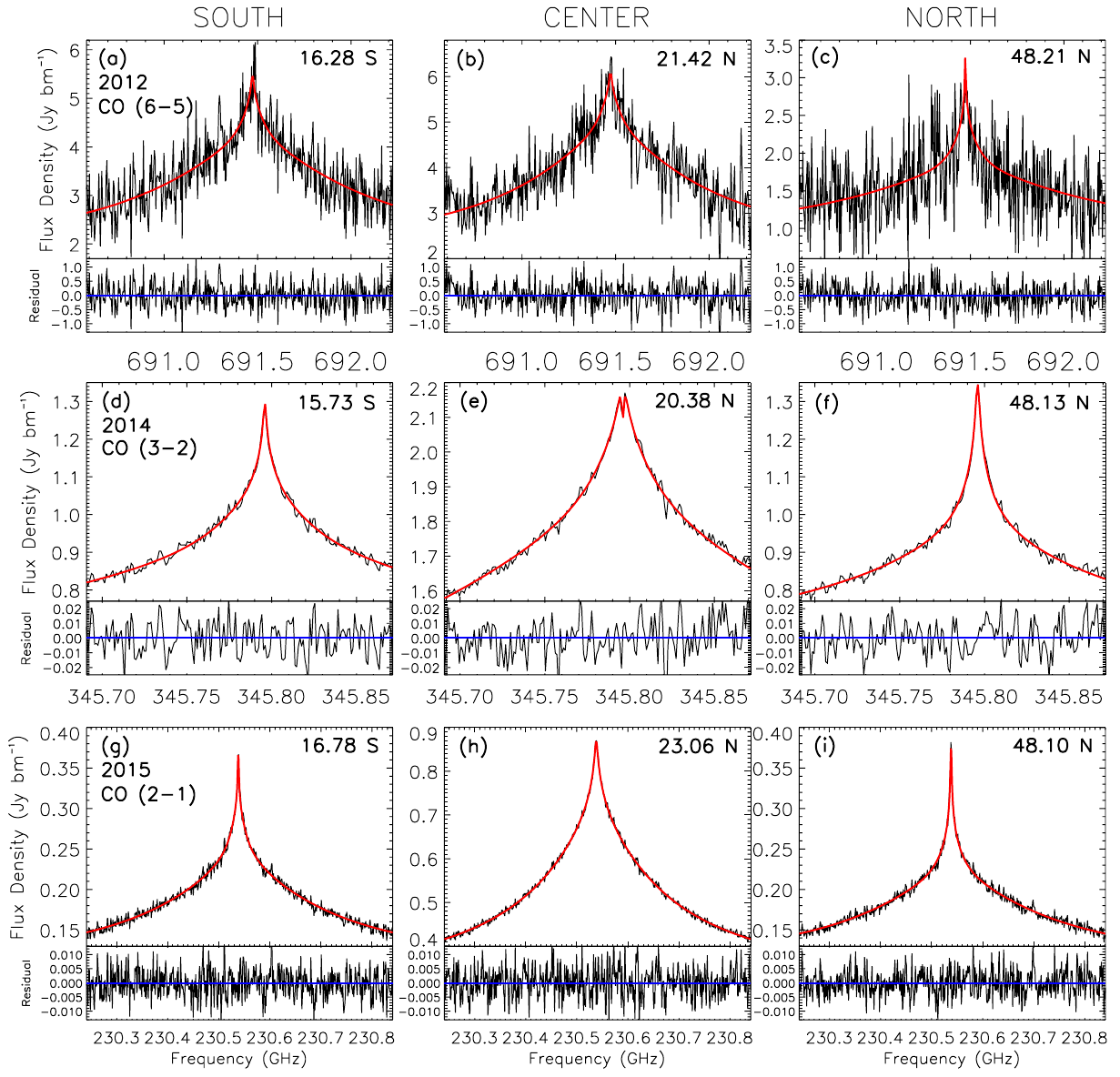}
\caption{ALMA spectra (black) and synthetic spectra generated by
    the NEMESIS radiative transfer code (red) for spatially resolved
    datasets from 2012 (a--c), 2014 (d--f), and 2015 (g--i). Mean
    latitudes for beam-footprint regions are shown, which correspond
    to regions in Fig. \ref{fig:beam}. Bottom panels show the residual
  flux as in Fig. \ref{fig:spec1}.}
\label{fig:spec2}
\end{figure}

The resulting disk-averaged synthetic spectra are
shown in Fig. \ref{fig:spec1}. Though our model atmosphere extends
from 0--1200 km, we found the temperature sensitivity of our
observations to be greatest between $\sim10^{2}-10^{-3}$ mbar
(approximately 50--530 km), as shown by contribution functions in
Fig. \ref{fig:cf}. We proceeded to use disk-averaged measurements as
\textit{a priori} temperature profiles to model spatially
resolved spectra, enabling spatial temperature profiles to be retrieved without the use
of corresponding \textit{Cassini} data in the stratosphere through
mesosphere. For the 2014 and
2015 datasets, \textit{Cassini} radio science and HASI data were
convolved with the ALMA restoring beam (in the locations specified in
Fig. \ref{fig:beam}) to produce
interpolated temperature profiles for use as \textit{a priori}
values in Titan's troposphere. This resulted in better fits to the continuum and
ensured that flux density scaling factors were within the errors of the
Butler-JPL-Horizons 2012 model for CO lines. Spatial
spectral models are shown in Fig. \ref{fig:spec2}. Finally, the
retrieved 2012 disk-averaged and spatial temperature profiles are shown
in Fig. \ref{fig:combo_1}a, with variations in 2013--2015
disk-averaged profiles
shown in \ref{fig:combo_1}b. Deviations from 2012 spatial temperature profiles
for 2014 and 2015 data are shown in Fig. \ref{fig:combo_1}c--e. All
retrieved temperature profiles are available to download
online ([Dataset] \cite{thelen_17b}\footnote{http://dx.doi.org/10.17632/xk3nkvz28b.1}).


\section{Discussion} \label{discussion}
\subsection{\textit{Cassini} Comparisons}
\indent To validate our model of Titan's atmosphere and the retrieved
temperature profiles presented in Fig. \ref{fig:combo_1}, it is useful
to compare these temperature measurements to those made by \textit{Cassini} in
similar latitude regions. Interpolating the HASI \parencite{fulchignoni_05} and \textit{Cassini} radio occultation
science measurements from the T27, T31, T46, and T57
flybys \parencite{schinder_12} at the latitudes of the
ALMA restoring beam (see Fig. \ref{fig:beam})
provides well constrained
temperatures from 7 distinct latitude regions on Titan's disk for
altitudes 0--100 km. Though these data contain temperature
measurements of Titan's stratosphere, they are not close enough in
time to warrant detailed comparisons, as stratospheric temperatures change on
much shorter timescales than those in the
troposphere \parencite{flasar_81, flasar_14}. Thus, for stratospheric temperature comparisons, we elected to extract
similar data from CIRS nadir maps taken during the T84, T98, T100, and
T112 flybys of Titan. These maps -- produced by modeling thermal
infrared spectra within the P and Q branches of the $\nu_4$ CH$_4$ band (1251--1311 cm$^{-1}$), as described in
detail in \textcite{achterberg_08} -- provide exceptional latitude
coverage ($2.5^\circ$ resolution) over Titan's disk during its transition into northern summer. Specific flybys were chosen to maximize
latitudinal coverage from temporally comparable measurements. These
flybys are listed in Table \ref{tab:cassini}, with the corresponding
data available online ([Dataset] \cite{achterberg_17}\footnote{http://dx.doi.org/10.17632/f3b9zj96tm.1}). We compare these two datasets to
retrieved disk-averaged and spatially resolved temperature profiles in
Fig. \ref{fig:combo_2}.


Temperature profiles obtained from ALMA retrievals are generally in good
agreement with both disk-averaged \textit{Cassini}/CIRS
measurements (Fig. \ref{fig:combo_2}a) and those from similar latitudinal
regions (Fig. \ref{fig:combo_2}b--d). Discrepancies between CIRS and ALMA
measurements in the stratosphere (100--300 km) are mostly less than 5
K. Uncertainties in our ALMA temperature retrievals range between \textpm2--5 K in
this altitude range, while CIRS measurements are accurate to \textless1
K \parencite{achterberg_08}. We find slightly warmer (average 1--2 K,
maximum 5 K)
disk-averaged stratospheric temperatures than CIRS for each year near
1 mbar (200 km),
except 2014. Northern and central spatial temperature measurements for 2012
(Fig. \ref{fig:combo_2}b--c solid lines) are
generally cooler than CIRS by up to $\sim$4 K, while southern
temperatures (Fig. \ref{fig:combo_2}d) are comparably warmer. These
variations largely lie within the retrieved temperature profile errors
(Fig. \ref{fig:combo_2}, dark gray regions) until below 5 mbar, where CIRS
profiles generally are less sensitive and relax back to \textit{a
  priori} values. This sharp cutoff is present in all disk-averaged
profiles and spatial 2014 and 2015
variations as well, resulting in 5--10 K warmer ALMA retrievals near
10 mbar. Northern and central temperature profiles from 2014 are warmer than CIRS
measurements by up to 4 K between 0.5--1 mbar. Spatial retrievals of
2015 data yield the largest disparities, particularly for the southern and central regions, which
were both warmer than CIRS by up to 5 K near 1 mbar. 

\begin{figure}[H]
\centering
\includegraphics[scale=0.45]{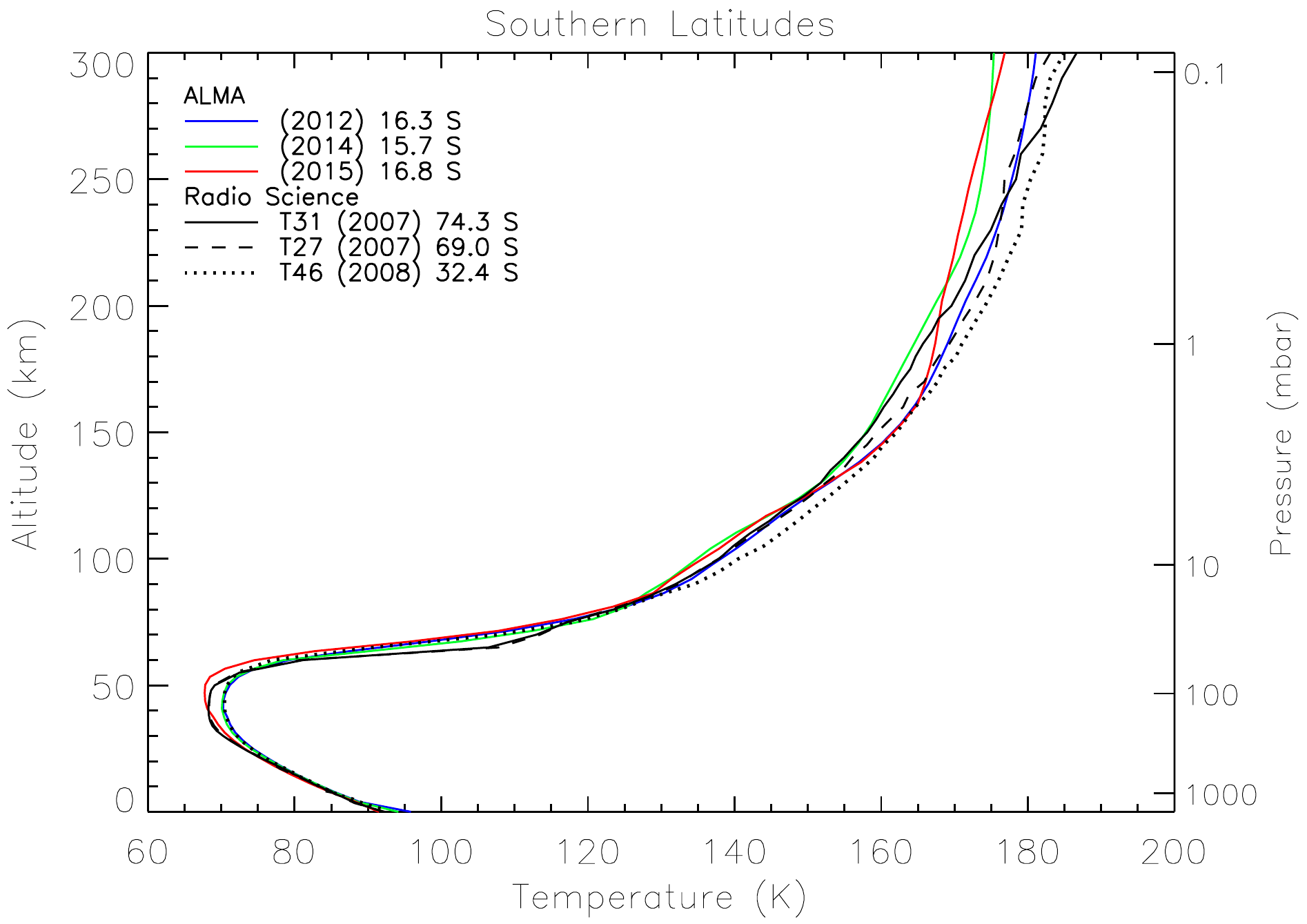}
\includegraphics[scale=0.45]{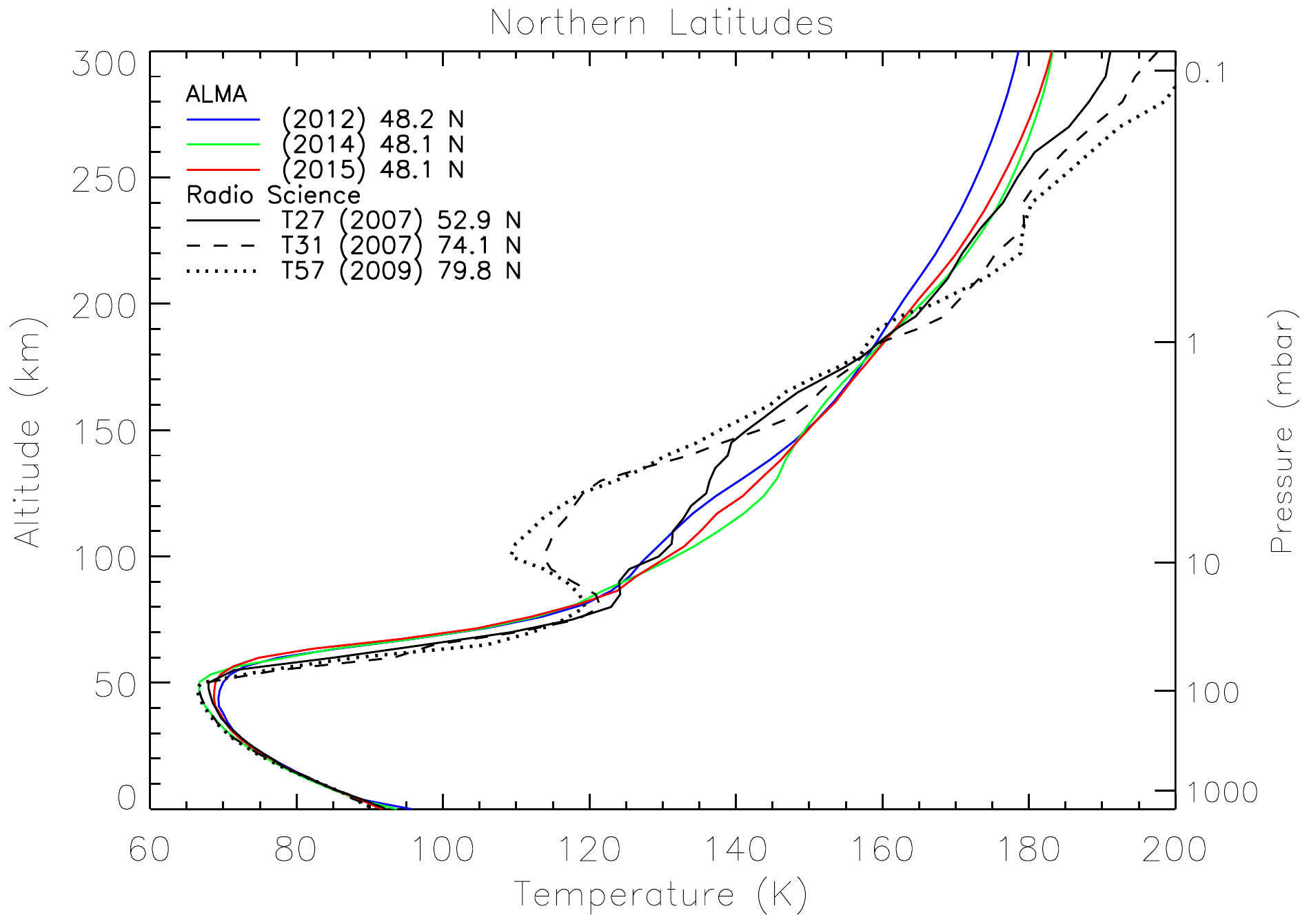}
\caption{Comparison of spatial ALMA retrievals from 2012 (blue lines), 2014
    (green), and 2015 (red) to radio occulation results
  (\cite{schinder_12}, Table \ref{tab:cassini}) at southern (left) and northern (right)
  latitudes (black lines). Radio science measurements were not convolved with ALMA
  beam-footprints to show seasonal variations in the lower atmosphere
  from 2007--2015, particularly above 100 km.}
\label{fig:tempcomp_radio}
\end{figure}

We obtain temperature profiles that are cooler than interpolated radio occultation
measurements by up to 10 K for all disk-averaged and spatially
resolved measurements between 50--100 km (Fig. \ref{fig:combo_2}e--h). At lower altitudes, ALMA
retrievals are no longer sensitive to temperature
(Fig. \ref{fig:cf}), and thus adhere to the \textit{a priori} profile. Above 100 km, temperatures retrieved through
radio occultations have uncertainties on the order 1--10 K \parencite{schinder_12}, and we defer to CIRS nadir measurements
which have lower errors and are from more recent \textit{Cassini} observations.
To illustrate the
potential discrepancies between our presented ALMA retrievals and interpolated radio
occultation data from 2007--2009 as a result of seasonal changes, particularly above the tropopause, we compare our retrieved
temperature profiles to those published by \textcite{schinder_12} up
to 300 km in Fig. \ref{fig:tempcomp_radio}. From 0--300 km, southern profiles generally agree with those obtained by radio occultations despite
considerable latitudinal and temporal differences. However, in northern latitudes, we do not observe previous
instabilities as observed by \textit{Cassini}, potentially caused by cloud formation or enhanced
photochemical production during northern winter \parencite{schinder_12,
  flasar_14}.


Most CIRS temperature variations above and below 1 mbar are 
within the range of ALMA retrieval errors, as are those for the radio
occultation data below 30 mbar ($\sim$65 km); however, larger
disparities arise for the 2014 and 2015 datasets in both the
disk-averaged and spatially resolved cases than for 2012 and 2013
measurements in both regimes. This is most likely due to
the vast improvement of ALMA data between 2012 and 2015
observations due to the expanded interferometer (see `$\#$ of
Antennae' in Table \ref{tab:obs}), resulting in increased coverage of
the \textit{u-v} plane and higher S/N spectra (see
Fig. \ref{fig:spec1}). This makes minor flux calibration issues, solved
by applying a small multiplicative scaling factor (discussed in
Section \ref{results}), result
in greater variations between retrieved temperature profiles and
\textit{Cassini} measurements (see Fig. \ref{fig:scale}). 
Many uniform scaling factors applied to the spectral radiance (of
order $\sim5\%$)
resulted in temperature retrievals with much larger deviations from
\textit{Cassini} data, often yielding cold (\textless65 K)
tropopause and hot (\textgreater200 K) stratospheric
temperatures. Generally, these factors are larger for high spatial resolution
spectra due to the large temperature gradients that exist at high
latitudes \parencite{schinder_12, coustenis_16} and are not
accounted for in the flux calibration model. These systematic errors
provide a motivation for improving the ALMA flux calibration model of
Titan to incorporate the effects of latitudinal and seasonal
variations in tropospheric and stratospheric temperatures,
respectively, which can produce large uncertainties in
pressure-broadened lines in Titan's atmosphere (such as CO and HCN).
For disk-averaged spectra, however, continuum measurements are close enough to
our model spectra (within $\sim2\%$) to provide a reliable
calibration source for other ALMA science targets; Titan continuum
windows are often used to set the flux density for science objects and
spectral regions containing CO, HCN, and other strong lines on Titan itself.

Despite the aforementioned uncertainties, we find that applying a
two-sample Kolmogorov-Smirnov (KS) test to the temperature profiles
presented here with respect to those from \textit{Cassini}/CIRS and radio
occultation science
reveals a high degree of correlation, and that variations --
particularly those $\leq$5 K, often within the retrieved errors --
are not statistically significant. These statistics are detailed in
Table \ref{tab:chisq}, with the KS test statistic (D), and corresponding significance
level ($\alpha$), shown for all retrieved profiles compared to \textit{Cassini}/CIRS and
interpolated radio occultation profiles at altitudes \textless100 km. Though variations in retrieved ALMA and radio science measurements do
not seem statistically significant (i.e. $\alpha$\textgreater0.10),
general KS significance levels are lower than for CIRS comparisons, and lower
stratospheric ($\geq$60 km) temperature deviations are often larger
than the retrieved error for ALMA observations (Fig. \ref{fig:combo_2}e--h). 

\subsection{Spatial and Temporal Variations}
\begin{figure}[H]
\centering
\includegraphics[scale=1.]{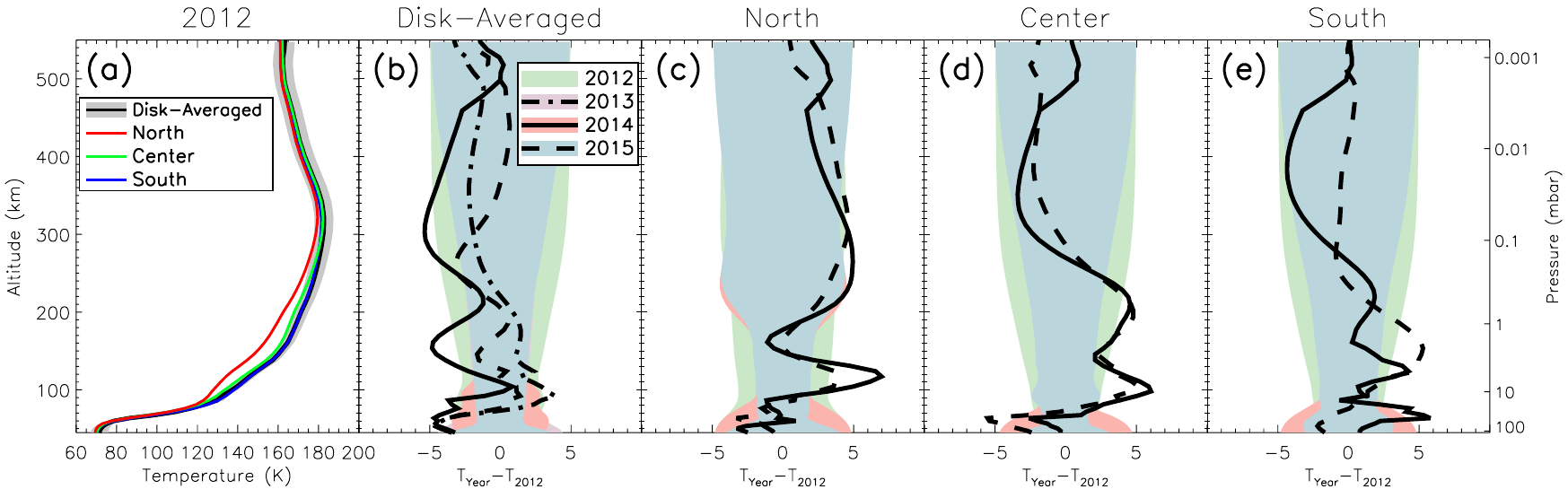}
\caption{(a) Retrieved temperature profiles for 2012 disk-averaged
    (black line) and spatial spectra (blue, green, and red) fit by the
    NEMESIS radiative transfer code shown in
    Fig. \ref{fig:spec1} and \ref{fig:spec2}, respectively. The 1-$\sigma$ error
    envelope of the disk-averaged retrieval is shown in gray. Temperature
    profiles are plotted where the CO retrievals are most sensitive,
    as shown in Fig. \ref{fig:cf}. (b) Comparison of temperature
    difference in disk-averaged
    profiles from 2013 (dash dot), 2014 (solid),
    and 2015 (dashed) with respect to 2012 from panel a. Error envelopes for 2012, 2013, 2014, and 2015
  datasets are shown (green, purple, red, and blue, respectively). (c--e) Variations in
spatial profiles for 2014 and 2015 data compared to the 2012 profiles
from panel a; 2012, 2014, and 2015 error envelopes are shown.}
\label{fig:combo_1}
\end{figure}

\begin{figure}[H]
\centering
\includegraphics[scale=1.]{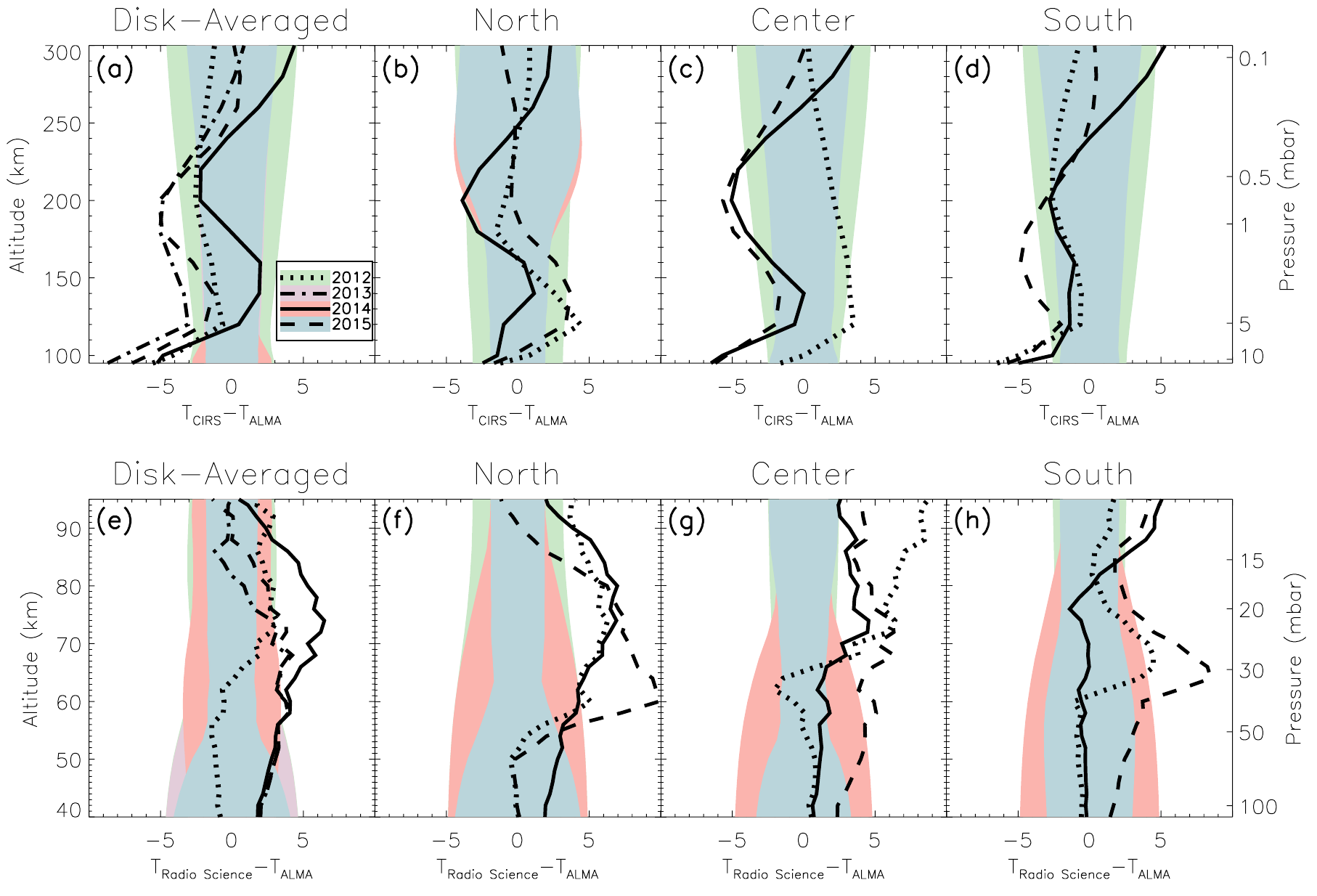}
\caption{(a--d) Difference between CIRS
    nadir maps and ALMA retrievals for disk-averaged (panel a) and spatially resolved profiles
    (panels b--d) from 2012 (dotted lines), 2013, (dash dot), 2014
    (solid), and 2015 (dashed) datasets. 1-$\sigma$ error envelopes
    are shown in color as in Fig. \ref{fig:combo_1}. CIRS nadir data
    were convolved with ALMA beam-footprints (Fig. \ref{fig:beam}) and
  are sensitive between 0.1--10 mbar ($\sim$100-300 km). (e--h)
  Deviations of ALMA retrievals from Cassini radio occultation
  science (\cite{schinder_12}) and HASI
  data (\cite{fulchignoni_05}) interpolated at ALMA beam-footprint locations. Radio science data are accurate to
  within 1 K up to $\sim$10 mbar (100 km).}
\label{fig:combo_2}
\end{figure}

\begin{figure}[H]
\centering
\includegraphics[scale=1.]{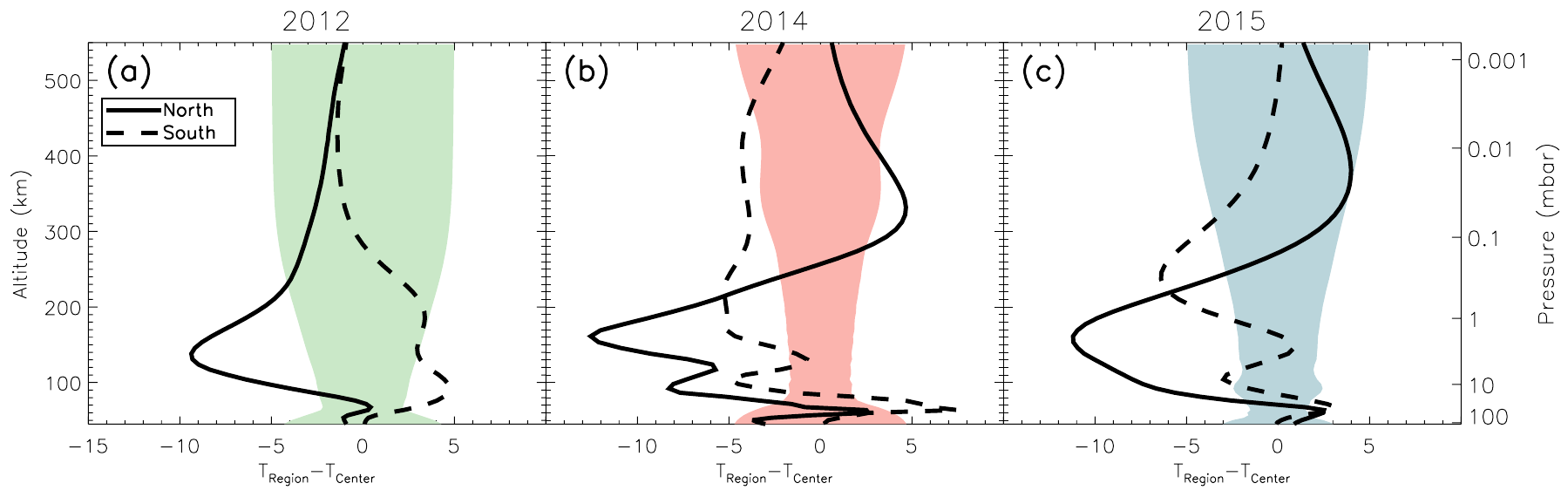}
\caption{Year by year comparisons of differences between North (solid lines) and South
    (dashed) temperature profiles with temperatures extracted from
    Center regions. 1-$\sigma$ error
  envelopes for Center profiles are shown in color.}
\label{fig:combo_3}
\end{figure}

We observe little variation in disk-averaged temperature profiles from
2012--2015, as shown in Fig. \ref{fig:combo_1}b. These
measurements agree with previous disk-averaged temperature
profiles obtained with ALMA \parencite{serigano_16} and with CIRS nadir measurements
(Fig. \ref{fig:combo_2}a), which both show small dispersion between
retrieved disk-averaged temperatures. Temporal and spatial variations become
apparent, however, when comparing individual regions on Titan's
disk. Variations by region are shown in Fig. \ref{fig:combo_1}c--e, and
yearly comparisons of north and south regions to central profiles are shown
in Fig. \ref{fig:combo_3}. 


Northern regions show general warming over all three
years throughout the stratosphere (particularly from 100--300 km), as the northern hemisphere receives
higher insolation during the transition into northern summer. Stratopause ($\sim$310--330 km) temperatures for both
2014 and 2015 are warmer than measured in 2012 by about 5 K (Fig. \ref{fig:combo_1}c), in good
agreement with \textcite{coustenis_16}. The northern stratosphere from
$\sim$80--250 km
generally remains cooler than the central latitudes by 10 K (Fig. \ref{fig:combo_3}),
consistent with CIRS limb and nadir
observations throughout the \textit{Cassini}
mission \parencite{vinatier_15,coustenis_16}. The stratopause,
however, becomes warmer than central and southern profiles by 5 and 10
K, respectively, during 2014 and
2015. 

Southern temperatures rise in
the stratosphere from 2012--2015 by up to 5 K, particularly between
1--10 mbar, and throughout the strato- and mesosphere from
2014--2015 by a similar amount. This is explained by increased downwelling of
Titan's large Hadley-type circulation cell, which may warm the upper
stratosphere and mesosphere (\textgreater300 km) of the winter
pole substantially. This has been observed in both \textit{Cassini}
observations \parencite{teanby_12, vinatier_15} and general circulation models
(GCM) of Titan \parencite{newman_11, lebonnois_12}. Southern profiles
remain warmer than those at high northern latitudes throughout
2012--2015 in the lower stratosphere by 5--15 K, and cooler than the central
profiles in 2014--2015 by $\sim$5 K. These results are explained by
the viewing geometry of Titan as seen by ALMA, as the southern latitudes we observe are
relatively low. Temperature profiles in the south will not be truly indicative of the
winter pole, where the stratosphere should be quite
cold \parencite{coustenis_16}; indeed, deviations from the center -- which is less effected
by seasonal variations and reduced insolation -- are much less
pronounced in the south than the north. The central region follows a
similar trend to the north in the lower stratosphere, though reduced
in magnitude. The upper atmosphere cools in 2014 and rises again in 2015
similar to southern profiles, yet these changes are well within the
retrieved errors.

We observe general cooling of the lower stratosphere ($\leq$100 km) in all three regions over
2012--2015 within the retrieved errors, with the exception of the south from 2012--2014 and the
center from 2012--2015; here we
observe heating above the tropopause by up to 5 K in the south and
cooling by the same amount in the center
(Fig. \ref{fig:combo_1}d--e). While seasonal changes at these
altitudes are dampened compared to the stratosphere, we observe
consistent latitudinal variations (Fig. \ref{fig:combo_3}) below 100
km as observed in \textit{Cassini} radio occultation measurements \parencite{schinder_12}. Southern profiles are
consistently warmer than central measurements by up to 8 K,
particularly near 60 km. Northern profiles tend to match central
temperatures (within the errors) at these altitudes. Considerable variability does exist, however, between
northern profiles for all three years near 10 mbar (100 km), where seasonal effects may begin to
manifest in atmospheric temperatures over smaller timescales. We observe the largest temporal temperature
variation of all profiles presented here, 7 K, from 2012--2014 at
$\sim$ 120 km (Fig. \ref{fig:combo_1}c), despite remaining colder than central and southern
temperatures by 10--15 K (Fig. \ref{fig:combo_3}). However, this increase in
temperature subsides from 2014--2015. While these variations are not
as significant as those observed by \textit{Cassini} at higher
latitudes \parencite{coustenis_16}, this region is of particular
interest for future ALMA studies. Submillimeter CO emission is
particularly sensitive to temperatures at these altitudes
(Fig. \ref{fig:cf}), providing insight into temperature variations below
the CIRS sensitivity range and where radio science uncertainties
become large. Further, these altitudes are high enough to not be significantly impacted by
uncertainties in flux calibration scaling factors
or models of the continuum (formed near the
tropopause), which often manifest as large variations in the
stratopause and tropopause (Fig. \ref{fig:scale}c).

Temperatures in the mesosphere (altitudes $\geq$ 350 km) cool in central
and southern regions from 2012--2014 by up to 5 K, and
rise in the north by similar amounts. However, from 2014 to 2015,
these variations are often reduced substantially or reversed, as in
the center and south. As the radiative
dampening time of Titan's upper atmosphere is less than a year, mesospheric
temperatures become highly variable \parencite{achterberg_11,
  teanby_12, flasar_14}, though these
results are consistent with an adiabatic cooling of the summer pole by
5 K in GCM studies \parencite{newman_11}. This provides further
motivation for increased observation of Titan with ALMA
in the coming years -- preferably in intervals of less than 1 year -- with
high spatial resolution.

While general spatial trends tend to agree with those found in
GCMs and \textit{Cassini} measurements, it should be noted that our temperature estimates
comprise a weighted average of latitudes (see Fig. \ref{fig:beam}),
and thus complex latitudinal variations in temperature structure are
accordingly dampened. However, the spatial variations present in
measurements shown here are still substantial across Titan during each
year -- up to 15 K from northern to southern latitudes
(Fig. \ref{fig:combo_3}). This reveals the importance of utilizing
correct temperature profiles in future modeling efforts of spatially resolved ALMA datasets, particularly for
retrievals of chemical abundance at high latitudes. Though ALMA observations of Titan will not have
the same degree of temporal cadence and high latitude resolution as \textit{Cassini},
ALMA's increasing spatial resolution (to \textless20 mas) in the coming years will allow
us to produce maps of Titan's atmospheric temperature and chemical
abundance, similar to those obtained with \textit{Cassini}/CIRS
measurements.


\section{Conclusions}
Retrieved temperature profiles from CO emission lines in spatially
resolved ALMA datasets of
Titan provide an avenue to observe distinct temporal and
latitudinal variations between $\sim$50--500 km, despite the constraints present in ALMA flux calibration
observations due to
viewing geometry and spatial resolution. These
measurements are particularly sensitive between 60--300 km, providing a
highly complementary dataset to \textit{Cassini} observations in the IR and
through radio occultations. We observe a warming of Titan's
stratopause ($\sim$320 km) in northern latitudes by up to 5 K from
2012--2015, and an increase in temperature in the lower
stratosphere by an equal amount in low southern latitudes  -- both indicators of increased insolation in
the northern summer and of increased downwelling in the winter
hemisphere due to Titan's global circulation cell. We observe a surprising
increase in temperature of the lower stratosphere in northern latitudes from 2012--2014, and
generally warmer temperatures in the north and colder
temperatures in the south than those measured by radio
occultations at $\geq$80 km. While temporal
variations are often within the retrieved errors, limited by the short
integration times of flux calibration datasets and large beam sizes,
larger temperature variations are present between spatial
regions. Here, we observe latitudinal temperature differences up to 15 K between northern latitudes and regions near
the equator. 

The validation of these measurements by
\textit{Cassini} observations is crucial to the continuation of Titan studies
with ground-based facilities. Our retrieved temperature profiles are in good agreement with contemporary
\textit{Cassini}/CIRS nadir maps -- with deviations mostly \textless5
K -- and previous GCM studies. We find that deviations from
\textit{Cassini}/CIRS observations between 100--300 km and radio science
measurements below 100 km are not statistically significant
(Fig. \ref{tab:chisq}) and are largely within the 1-$\sigma$ retrieval
errors. While these data are representative of flux calibration
observations taken before 2016, ALMA's longest baselines (16 km) will allow
for observations with beam sizes \textless20 mas, resulting in many (10--100) resolution elements across Titan's
disk. The high sensitivity of the completed array may
allow for the detection and mapping of additional atmospheric species during
dedicated observations with longer integration times. Thus, not
only do these data allow us to confidently monitor
Titan's atmospheric dynamics beyond the end of the \textit{Cassini} mission,
they also greatly improve our ability to perform spatially resolved retrievals of
chemical abundance of the many trace constituents present in Titan's
atmosphere that are observable with ALMA. 

\section{Acknowledgments}
This research was supported by NASA’s Office of Education and the NASA Minority University Research and Education Project ASTAR/JGFP Grant $\#$NNX15AU59H. 

Additional funding was provided by the NRAO Student Observing Support award $\#$SOSPA3-012.
	
This paper makes use of the following ALMA data:
ADS/JAO.ALMA$\#$2012.1.00317.S, 2012.1.00688.S, 2011.0.00724.S, and 2012.1.00501.S. ALMA is a partnership of ESO (representing its member states), NSF (USA) and NINS (Japan), together with NRC (Canada) and NSC and ASIAA (Taiwan) and KASI (Republic of Korea), in cooperation with the Republic of Chile. The Joint ALMA Observatory is operated by ESO, AUI/NRAO and NAOJ. The National Radio Astronomy Observatory is a facility of the National Science Foundation operated under cooperative agreement by Associated Universities, Inc.



\printbibliography[title={References}]


\begin{table}[H]\footnotesize
\begin{center}
\caption[]{\em{Observational Parameters}}
\begin{tabular}{cccccccc}
\toprule
Transition & Observation & Rest Freq. & Integration & $\#$ of & Spectral &
  Beam & Project \\
 & Date & (GHz) & Time (s) & Antennae & Res. (kHz) & Size$^a$ & ID \\
\midrule
CO (6--5) & 05 Jun 2012 & 691.473 & 236 & 21 & 976 & 0.35$''$ $\times$ 0.28$''$ &
                                                             2011.0.00724.S
  \\
CO (2--1) & 14 Dec 2013 & 230.538 & 157 & 28 & 122 & 1.54$''$ $\times$ 0.77$''$ &
                                                             2012.1.00688.S
  \\
CO (3--2) & 15 Jun 2014 & 345.796 & 157 & 36 & 976 & 0.39$''$ $\times$ 0.34$''$ &
                                                             2012.1.00501.S
  \\
CO (2--1) & 27 Jun 2015 & 230.538 & 157 & 42 & 976 & 0.37$''$ $\times$ 0.32$''$ &
                                                             2012.1.00317.S
  \\
\bottomrule
\label{tab:obs}
\end{tabular}
\end{center}
{\small{{\bf{Notes:}} $^a$FWHM of the Gaussian restoring beam}}
\end{table}

\begin{table}[H]\footnotesize
\begin{center}
\caption[]{\em{Cassini Data}}
\begin{tabular}{ccccc}
\toprule
 Date & Titan & Latitude & Altitude & L$_S$ \\
 & Flyby & Coverage$^a$ & Range (km) & ($^\circ$) \\
\midrule
 CIRS\\
\midrule
 06 Jun 2012 & T84 & 80.0$^\circ$ S -- 72.5$^\circ$ N & 100--300 & 34.05 \\
02 Feb 2014 & T98 & 90.0$^\circ$ S -- 87.5$^\circ$ N & 100--300 & 53.16 \\
07 Apr 2014 & T100 & 90.0$^\circ$ S -- 90.0$^\circ$ N & 100--300 & 55.15 \\
07 Jul 2015 & T112 & 67.5$^\circ$ S -- 72.5$^\circ$ N & 100--300 & 69.17 \\
\midrule
 Radio Occultation Science\\
\midrule
 26 Mar 2007 & T27 & 69.0$^\circ$ S, 52.9$^\circ$ N & 0--300 & 329.52 \\
28 May 2007 & T31 & 74.3$^\circ$ S, 74.1$^\circ$ N & 0--300 & 331.78 \\
03 Nov 2008 & T46 & 32.4$^\circ$ S, -- & 0--300 & 350.32 \\
22 Jun 2009 & T57 & 79.8$^\circ$ N, -- & 0--300 & 358.29 \\
\bottomrule
\label{tab:cassini}
\end{tabular}
\end{center}
{\small{{\bf{Notes:}} $^a$Latitude coverage for Cassini CIRS nadir
    measurements are continuous, with 2.5$^\circ$ latitude bins; radio occultation science latitudes
    are given for ingress and egress observations (near the surface), respectively.}}
\end{table}

\begin{table}[H]\footnotesize
\begin{center}
\caption[]{\em{Statistical Tests}}
\begin{tabular}{c@{\qquad}cc@{\qquad}cc}
  \toprule
  \multirow{2}{*}{Measurement} & \multicolumn{2}{c}{Radio Science} & \multicolumn{2}{c}{CIRS} \\
  \cline{2-5}\\[-2.5ex]
   & D$^a$ & $\alpha$$^b$ & D & $\alpha$ \\     
  \midrule
  Disk-Averaged \\
  \midrule
  2012 & 0.12 & 0.99 & 0.18 & 0.99 \\      
  2013 & 0.16 & 0.88 & 0.18 & 0.99 \\
  2014 & 0.16 & 0.88 & 0.27 & 0.74 \\      
  2015 & 0.16 & 0.88 & 0.18 & 0.99 \\
  \midrule
  Spatially Resolved \\
  \midrule
  2012 (South)  & 0.08 & 0.99 & 0.12 & 0.99 \\      
  2012 (Center) & 0.24 & 0.41 & 0.09 & 1.00 \\ {\smallskip}     
  2012 (North)  & 0.24 & 0.41 & 0.09 & 1.00\\
  
  2014 (South)  & 0.20 & 0.65 & 0.27 & 0.74 \\      
  2014 (Center) & 0.12 & 0.99 & 0.18 & 0.99 \\ {\smallskip}     
  2014 (North)  & 0.20 & 0.65 & 0.27 & 0.74\\
  
  2015 (South)  & 0.16 & 0.88 & 0.18 & 0.99 \\      
  2015 (Center) & 0.20 & 0.65 & 0.18 & 0.99 \\   
  2015 (North)  & 0.12 & 0.99 & 0.09 & 1.00 \\
  \bottomrule
\label{tab:chisq}
\end{tabular}
\end{center}
{\small{{\bf{Notes:}} $^a$KS test
    statistic. $^b$Significance level of KS test statistic.}}
\end{table}

\end{document}